\documentclass[12pt]{iopart}
\usepackage{graphicx}  

\usepackage{iopams}  
\begin{document}

\title[Comments on the 2001 run of the 
EXPLORER/NAUTILUS ...]{Comments on the 2001 run of the 
EXPLORER/NAUTILUS gravitational wave experiment}

\author{
P Astone\dag, D Babusci\ddag, M Bassan\S, P Bonifazi$\|$,
P Carelli\P, \\G Cavallari$^+$, E Coccia\S, C Cosmelli$^*$, 
S D'Antonio\S, \\V Fafone\ddag, S Frasca$^*$, G Giordano\ddag, 
A Marini\ddag, Y Minenkov\S, \\I Modena\S, G Modestino\ddag, 
A Moleti\S, G V Pallottino$^*$, \\G Pizzella$\sharp$, L Quintieri\ddag, 
A Rocchi\S, F Ronga\ddag, R Terenzi$\|$, \\G Torrioli\dag\dag 
and M. Visco$\|$}

\address{\dag\ Istituto Nazionale di Fisica Nucleare INFN, Rome, Italy}
\address{\ddag\ Istituto Nazionale di Fisica Nucleare INFN, Frascati, Italy}
\address{\S\ University of Rome ``Tor Vergata" and INFN, Rome II, Italy}
\address{$\|$\ IFSI-CNR and INFN, Rome, Italy}
\address{\P\ University of L'Aquila and INFN, Rome II, Italy}
\address{$^+$\ CERN, Geneva, Switzerland}
\address{$^*$\ University of Rome ``La Sapienza" and INFN, Rome, Italy}
\address{$\sharp$\ University of Rome ``Tor Vergata" and INFN, Frascati, Italy}
\address{\dag\dag\ IFN-CNR and INFN, Rome II, Italy}

\begin{abstract}
The recently published analysis of the coincidences between the EXPLORER 
and NAUTILUS gravitational wave detectors in the year 2001 \cite{nostro} 
has drawn some criticism \cite{finn}. We do not hold with these objections, 
even if we agree that no claim can be made with our data. The paper we 
published reports data of unprecedented quality and sets a new procedure 
for the coincidence search, which can be repeated again by us and by 
other groups in order to search for signature of possible signals. 
About the reported coincidence excess, we remark that it is not destined 
to remain an intriguing observation for long: it will be confirmed or denied 
soon by interferometers and bars operating at their expected sensitivity.
\end{abstract}

\section{Comments on the experimental results}

At present, the 2001 scientific run performed by our gravitational wave (GW) 
detectors EXPLORER and NAUTILUS \cite{nostro} constitutes the most sensitive 
experiment for the detection of GW bursts ever carried out. When we analyzed 
the data collected, we became convinced of the importance of communicating 
that unprecedented sensitivities were reached by our detectors and that the 
use of powerful tools in analysis of the data (energy consistency of the 
events and sidereal time analysis) led to a coincidence excess centered at 
sidereal hour 4, which was defined in the paper ``an interesting indication".

Criticisms to this analysis have been recently published \cite{finn}. A first 
criticism regards the mere existence of a coincidence excess. It is stated that 
the potential significance of the peak centered at sidereal hour 4 must be 
``diluted" because we did not declare before the analysis  (i.e. ``a priori") 
that we were searching for sources giving excess in that sidereal hour (i.e. 
galactic sources). In other words, the significance of the peak was recognized 
``a posteriori". As a consequence, the probability should be decreased of 
roughly a factor 24 because {\em a priori} the peak could have been found in any 
of the 24 sidereal hours. 

To be quantitative about what we called an ``interesting indication", let us 
report (see  Table 1) the Poisson probabilities of the observed number of 
coincidences with respect to the corresponding number of accidentals in 
intervals of increasing duration centered at sidereal time 4.

\begin{table}
\caption{Number of observed coincidences $n_c$ and integral number of accidentals 
$\bar{n}$ for various sidereal time intervals around hour 4. The last column gives 
the Poisson probability that a background fluctuation produces $n_c$ or more counts.}
\begin{indented}
\item[]\begin{tabular}{@{}llll}
\br
 Sidereal time period (h) & $n_c$ & $\bar{n}$ & P   \\
\mr
 4 & 4  & 0.92 & 0.0145 \\
 3 - 5 & 7  & 1.69 & 0.0018 \\
 2 - 6 & 8  & 3.45 & 0.025  \\
 1 - 7 & 10 & 5.01 & 0.032 \\
 0 - 8 & 13 & 6.2  & 0.011  \\
\br
\end{tabular}
\end{indented}
\end{table}

These probabilities are, at most, at the level of a few percent. Instead, going up 
to the entire 24 hour period, the overall number of coincidences (31) with respect 
to the accidentals (25) gives a Poisson probability of 0.14.

We are well aware that extraordinary claims need extraordinary evidence. By simply 
looking at these Poisson probabilities, we see that the coincidence excess we 
reported was certainly not strong enough to claim a detection, and in fact no such 
claim was made in our paper. However, two remarkable facts need to be underlined and 
give sense to the words ``interesting indication": 

\begin{itemize}
\item the peak is centered at a physically significant sidereal hour, corresponding 
to the most favorable orientation of the detectors with respect to sources in the 
Milky Way. The Galaxy is certainly the privileged place of the sources attainable 
by present GW detectors and we think that the experiment described in \cite{nostro} 
should be considered as based on the ``a priori" hypothesis of signals originating 
in the Galaxy. This was clearly indicated in our previous paper \cite{nostrovecchio} 
(pag. 248), ``No extragalactic GW signals should be detected with the present 
detectors. Therefore we shall focus our attention on possible sources located in 
the Galaxy." Recent work by Paturel and Baryshev \cite{Paturel}, quantitatively 
indicated the signature expected from galactic sources in bar detectors of different 
sensitivity, and in particular the presence of a peak centered at sidereal hour 4 
for EXPLORER and NAUTILUS. So we don't think that the reported Poisson probabilities 
should be further ``diluted" by a factor taking into account all the possible peak 
positions in sidereal time.

\item there is a strong energy correlation for the coincidence events during the 
sidereal hours interval 3 to 5 (see figure 8 of reference \cite{nostro}). For those 
events the output of the two detectors have a correlation coefficient of 0.96 and 
the slope of the linear regression line for NAUTILUS energy on EXPLORER energy is 
1.18, which, within the accuracy of the detector calibration, is in agreement with 
the hypothesis of having equal signals on the two parallel bar detectors. On the 
contrary, the events at the other sidereal hours exhibit a correlation coefficient 
of -0.19 (a very poor correlation, as expected for random values).
\end{itemize}

Another criticism is that, since a bar detector has a rather broad antenna pattern, 
one should expect a correspondingly broad peak in the sidereal time distribution 
of the coincidences, instead of the relatively narrow peak reported in \cite{nostro}. 
This argument is very weak, because, given a source distribution, the geometrical 
antenna pattern of a single detector is not representative of the distribution of 
the coincidences between two detectors. Each detector has its own detection efficiency, 
which depends on the detector noise, threshold, and on the signal level. The 
coincidence distribution depends on the product of the efficiencies of the two 
detectors, and may give a relatively narrow peak if, for instance, the signals are 
near the thresholds. The fact that the width of the peak depends on the signal level 
and detector threshold was already shown for a single detector in reference 
\cite{Paturel}, where the case of EXPLORER was explicitly considered.

For these reasons, we disagree with the objections to our paper, even if we agree that 
no claim of detection can be made with our data.

Last comment: we were told that some statements in our paper about the possible 
contribution of real GW signals to our data led the reader to think that we were 
claiming a discovery. We admit that, in spite of our conservative attitude, in a 
couple of phrases the possible presence of gw signals in our data  was considered.
On the other hand it should be recognized by any experimentalist that it is difficult 
to take data with detectors of unprecedented sensitivity, containing the reported 
indication, without at least being open to this possibility.

\section{The future}

In order to calculate the probability, or the degree of belief, of having observed 
real GW signals (and not simply the Poisson probability that a background fluctuation 
produces $n_c$ or more counts), one should introduce astrophysical models for the 
sources. We deliberately left out this step in our paper, for two reasons: because 
we wanted to keep separate the observations from the possible astrophysical 
interpretations and, moreover, because we thought that more data were needed to 
evaluate the (unexpected) possibility that many gw bursts at the level of 
$h \simeq 2 \cdot 10^{-18}$ are bathing the Earth. 

The exercise presented by Astone \cite{pia} at the workshop GWDAW 2002 outlines 
how we can proceed in the future: updating the degree of beliefs adding new 
pieces of information as more data become available. That paper is an important 
contribution to the debate (now starting) on how to evaluate and compare the results 
reported by  different gravitational wave experimental collaborations.

In conclusion, let us say that we are proud to have crossed the benchmark of 
``nihil obstat" upper limit on the strength of the wave: we find that the amplitudes 
and rate of our candidate events are not only compatible with existing experimental 
upper limits \cite{igec} but also they are permitted by theoretical upper limits on 
gravitational wave strengths based on cherished beliefs about the astrophysical 
structure of the Galaxy and about the physical laws governing gravitational 
radiation \cite{thorne}. This was not the case of the events observed by Weber with 
its first generation bars at room temperatures, which amplitudes were at least two 
orders of magnitude larger and which can be explained only by invoking unconventional 
hypotheses, like strong beaming by sources near the galactic center, or today being 
a very special time in the evolution of the Galaxy \cite{thorne}.

We remark that the indication we reported is not destined to remain an intriguing 
observation (and source of trouble ...) for long: the existence of bursts bathing 
the Earth at the level of $h \simeq 2 \cdot 10^{-18}$ and at a rate of many per year 
would be such an unexpected gift from nature that it will easily be confirmed or denied 
soon with interferometers and bars operating at their expected sensitivity.

\end{document}